# Time Divergence-Convergence Learning Scheme in Multi-Layer Dynamic Synapse Neural Networks

**Ali Yousefi**
Boston University, Department of Mathematics and Statistics
Harvard Medical School, Department of Neurosurgery
`ayousefi@mgh.harvard.edu`

**Theodore W. Berger**
University of Southern California
Professor of Biomedical Engineering

## Abstract

A new learning scheme called time divergence-convergence – TDC – is proposed for two-layer dynamic synapse neural networks – DSNN. DSNN is an artificial neural network model, in which the synaptic transmission is modeled by a dynamic process and the information between neurons are transmitted through spike timing. In TDC, the intra-layer neurons of a DSNN are trained to map input spike trains to a higher dimension of spike trains called a feature-domain, and the output neurons are trained to build the desired spike trains by processing the spike timing of intralayer neurons. The DSNN performance was examined in a jittered spike train classification task which shows more than 92% accuracy in classifying different spike trains. The DSNN performance is comparable with the recurrent multi-layer neural networks and surpasses a single-layer DSNN with a 22% margin. Synaptic dynamics have been proposed as the neural substrate for sub-second temporal processing; we can utilize TDC to train a DSNN to perform diverse forms of sub-second temporal processing. The TDC learning proposed here is scalable in terms of the synaptic adaptation of deeper layers of multi-layer DSNNs. The DSNN along with TDC learning proposed here can be used in to replicate the processing observed in neural circuitry and in pattern recognition tasks.

## 1 Introduction

Neural processing is the outcome of computation run in multiple layers of neurons and billions of synaptic connections [1]. The multiple layer topology of the cortical neural circuits is one of the brain's evolutionary mechanisms to build an excessive processing power. The brain has the capability to process signals that span over 12 orders of magnitude in time. Within this range, processes like cognitive and perception, which happen on a sub-second time scale, are directly connected to synaptic temporal dynamics. Short- and long-term plasticity – STP and LTP – are considered as principal synaptic characteristics shaping brain temporal processing. There is also experimental evidence that suggests STP and LTP of an individual synapse are simultaneously adjusted in response to external stimuli [2, 3, 4]. The synapse model implemented in DSNN augments both LTP and STP factors; in this research, we propose a learning mechanism to adjust both LTP and STP parameters of an individual synapse of a two-layer DSNN.

The key in training a dynamic synapse neural networks – DSNN – is to address the temporal processing run in different layers of the network. To build a proper learning for a multi-layer DSNN, we require to correctly predict the spike timing of intra-layer neurons in response to an input stimulus. The prediction problem is practically ill-defined in existing learning algorithms; in other words, these algorithms are generally developed without considering temporal processing in different layers of the network. Reservoir computing – RC – which has been proposed as a learning mechanism in the recurrent spiking neural networks, circumvents the prediction problem by ignoring the processing performed in network intra-layer neurons. The RC suggests massive recurrent networks in the deeper layers of the network, and it moves the learning to output layer [5, 6]. RC approaches assume that the recurrent network generates a high dimensional spike train enough to generate any desired spike-domain function in its output layer. In fact, the learning proposed in RC is only applicable to a single-layer DSNN, and it is inapplicable for learning a multi-layer DSNN. Stochastic sampling methods, including reinforcement learning – RL – rely on a random exploration of the network intra- and output layer neurons' spiking activity to estimate the optimal learning scheme [7,8,9]. The RL suffers from a slow convergence in multi-layer topology, and, similar to RC, disregards the temporal processes run in the network deeper layers. The other statistical approach is based on Maximum Likelihood (ML) method; using this framework, the network parameters are adjusted to maximize the likelihood of observed spiking activity [10, 11]. In practice, for large networks, since the number of parameters grows exponentially, the learning process using ML becomes complicated. Another powerful statistical approach is point-process framework; despite being a powerful tool for training SNNs, it requires further theories and methods to address learning in multi-layer SNNs [12, 13, 14]. The STDP, the spike-time formulation of the Hebbian learning, only justifies neural processing run on a millisecond timescale [15]. The STDP is an unsupervised learning rule that adjusts the local behavior of a neuron, and it is unclear how this local adaptation can coordinate network level functionality [16]. Gradient descent techniques, generally back-propagation and specifically Deep Learning methods, require measurement of the objective function gradient with respect to the free network parameters [17, 18, 19, 20]. For the multi-layer DSNN, calculation of the gradient becomes inaccurate given the discrete nature of spike generation process. The gradient calculation becomes even more complicated in multi-layer DSNNs, given the fact that the objective function spans time and spiking activity in the deeper layers is unobserved.

In time divergence-convergence – TDC, we take account the neural processing run in an individual synapse, and propose a learning algorithm which helps to avoid calculating complex gradient functions. In TDC, we recruit two different forms of spike-domain processing performed by a single layer DSNN to address learning in multi-layer DSNNs. The two forms of spike-domain processing are temporal selectivity – TS – in the intra-layer neurons, and temporal integration – TI – in the output layer of a DSNN. Using TS process, we build a high-dimensional representation of the input stimulus projected on spike timing of intra-layer neurons. The TI process combines the intra-layer spike timing to generate the desired spike train/s in the output layer. TDC predicts the optimal spike timing for each neuron of a DSNN; using this prediction, we only need to work with the learning insingle layer DSNNs.

The DSNN is a spike-domain processing engine; using TDC, we can train the DSNN to perform different forms of spike-domain tasks. These tasks might include replicating neural processing in a neural circuit, or application-oriented tasks including feature extraction and pattern recognition. The classification of jittered spike trains is a spike-domain benchmark problem has been suggested to assess how recurrent neural networks and specifically RC models perform temporal processing tasks [21]. Here, we demonstrate the DSNN performance in performing this task and compare it with other models.

### 1.1 Synapse FD-Model

The discrete representation of the FD model is defined by – for a more detailed explanation, you might check [22]:

$$F_{n+1} = (1 - \frac{1}{\tau_f} - \Delta F * 1_n^{ap}) * F_n + \frac{1}{\tau_f} * F_0 + \Delta F * 1_n^{ap} \quad (1a)$$

$$N_{n+1} = (1 - \frac{1}{\tau_r} - (\frac{1}{\tau_f} * F_0 + \Delta F) * 1_n^{ap}) * N_n$$

$$-\ (1 - \frac{1}{\tau_f} - \Delta F) * 1_n^{ap} * F_n * N_n + \frac{1}{\tau_r} \quad \text{(1b)}$$

$$G_{n+1} = (1 - \frac{1}{\tau_g}) * G_n + N_{max} * (\frac{1}{\tau_f} * F_0 + \Delta F) * 1_n^{ap} * N_n$$

$$+\ N_{max} * (1 - \frac{1}{\tau_f} - \Delta F) * 1_n^{ap} * F_n * N_n \quad \text{(1c)}$$

$$K_{n+1} = (1 - \frac{1}{\tau_k}) * K_n + N_{max} * (\frac{1}{\tau_f} * F_0 + \Delta F) * 1_n^{ap} * N_n$$

$$+\ N_{max} * (1 - \frac{1}{\tau_f} - \Delta F) * 1_n^{ap} * F_n * N_n \quad \text{(1d)}$$

$$V_n = G_n - K_n \quad \text{(1e)}$$

Variable $F_n$ describes the facilitation kinetics in response to an incoming spike - $1_n^{ap}$ at time $n$ becomes 1 for an incoming spike and zero otherwise. Variable $N_n$ represents the portion of release-ready vesicles. The facilitation increment and vesicle release dynamics are both a spike-driven process, where the spike time is defined by $1_n^{ap}$. The $F_n$ resting value is $F_0$, and the $N_n$ resting value is set to 1.0. Auxiliary variables $G_n$ and $K_n$ approximate an $\alpha$-synapse function representing the post-synaptic potential – PSP, defined by the variable $V_n$ – elicited by synaptic neurotransmitter release [23]. The parameter $\Delta F$ which controls the calcium influx in the pre-synaptic terminal of a synapse is the main biological factor modulating synaptic temporal dynamic; and the parameter $N_{max}$ is a post-synaptic factor representing number of neurotransmitter receptors or simply synaptic strength. The facilitation and recovery time constants are defined by $\tau_f$ and $\tau_r$; the $\tau_g$ and $\tau_k$ determine the PSP rise and fall time. Figure (1) shows the synaptic PSP in response to a 500 milliseconds spike train consisting of 10 consecutive spikes with 50 milliseconds inter-spike interval - ISI.

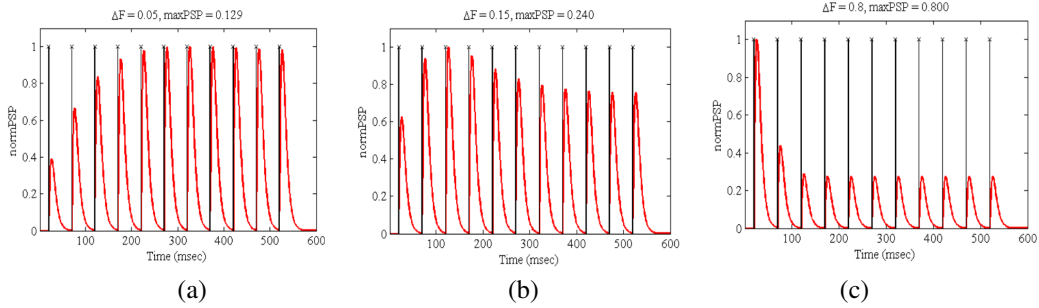

Figure 1: Normalized synaptic response to a 500-millisecond spike train with 50 milliseconds ISI. **(a)** A synapse with $\Delta F = 0.05$, **(b)** A synapse with $\Delta F = 0.15$, and **(c)** A synapse with $\Delta F = 0.8$. The $N_{max}$ parameter is set to 1.0; the black bars show spike timing. The PSP is normalized to its maximum response

The synapse model defined in equation set (1) is a nonlinear system; the non-linearity in the synapse model appears in the form of $F_n * N_n$ term, and it becomes linear if $F_n * N_n$ term is replaced by a surrogate variable. The linearization process will be completed by truncating higher-order non-linear terms of the synapse FD model, and then defining a set of new surrogate variables. The $F_n$ and $N_n$ are both positive values less than one; as a result, the higher-order terms of $F_n$ and $N_n$ tends to shrink to zero. Equation set (2) defines a linear state space model of the synaptic response - elements of the $A_s$, $A_t$, $B_s$ and $B_t$ matrices are continuous functions of the FD-model. $C$ is a vector with the same length of $X_k$; the value of its elements are defined using equation (1.e) considering new surrogate variables:

$$X_{n+1} = (A_s + 1_n^{ap} * A_t) * X_n + (B_s + 1_n^{ap} * B_t) \quad \text{(2a)}$$
$$V_n = C * X_n \quad \text{(2b)}$$

Equation (2) defines a general synapse model, which is a linear time-varying state-space model. For the modeling purpose, we assume that neuron integrates the PSP response of multiple synaptic connections plus a refractory process emerging from its previous spiking activity. The neuron fires a spike, whenever its membrane potential passes a firing threshold. Equation set (3) defines a neuron membrane potential dynamics and its spike generation criterion:

$$V_n = \sum_m V_n^m + \text{Ref}_n^o \quad (3a)$$
$$\text{Ref}_{n+1}^o = (1 - \Delta t/\tau_{ref}) * \text{Ref}_{n+1}^o + \text{Ref}_{\text{amp}} * 1_n^{ap,o} \quad (3b)$$
$$V_n > V_{thn} \geq V_{n-1} \Rightarrow 1_n^{ap,o} = 1 \quad (3c)$$

where, variable $V_n$ is the neuron membrane potential. Variable $V_n^m$ is the PSP of the $m^{th}$ synapse and variable $Ref_n^o$ corresponds to the refractory effect emerging by the neuron previous spiking activity – $1_n^{ap,o}$. $Ref_amp$ defines the change in the refractory signal on the neuron spike time. Equation set (2) and (3) formulate a linear-state representation of the neuron membrane dynamics with multiple synaptic connections; the linear state model can be applied for each neuron of a multi-layer DSNN, where the synaptic connections of each neuron are elicited by the spiking activity of their parent neurons.

**1.2 Temporal Processing in a Dynamic Synapse**

Spike-domain temporal processing is a mapping $f : S \to Q$, for which both $S$ and $Q$ are increasing sequences of $R$ – or $N$ for the discrete models as defined here – numbers – $S : 0 < s_1 < s_2 < \cdots < s_n$ and $Q : 0 < q_1 < q_2 < \cdots < q_m$. These numbers define spike timing for $S$ and $Q$ sequences.

A single synapse is capable of performing uni-modal temporal selectivity, in which the $Q$ sequence consists of any successive subset of the $S$ - with a possible small causal delay. Free parameters of a synapse - $\Delta F$ and $N_{max}$ or their equivalence in equation (2) - are mutually adjusted to perform different forms of the temporal selectivity process. Figure (2-b) shows different forms of the temporal selectivity, and figure (2-c) displays the optimum parameter set for the temporal selectivity tasks defined for a 500 milliseconds spike train with 50 milliseconds ISIs.

The TS and its variants extract different features of the input stimuli which can be part of further neural processing, decision making or classification. For example, the TS process is capable of building a sparse representation of the input spike timing; where, the input spike timing is encoded in the spike timing of multiple neurons, each of which fires in a pre-defined time window.

Dynamic synapse capability in performing the TS process degrades in prolonged spiking activities. For instance, the TS of the $10^{th}$ spike has a very limited choice of synaptic parameters - the tail of the TS curve in figure (2.c), which makes its synaptic response sensitive to a possible jitter in input ISIs. A single synapse is also incapable of performing bimodal temporal selection – B-TS - or a more intricate temporal processing. Note that, this is independent of possible variability in the impinging input, which might lead to a grouped spiking activity in the synapse output. Either successive layers of synaptic connections or a recurrent connection is required to boost the temporal processing capability of a DSNN. The TDC learning framework aims to address the synaptic adaptation in a two-layer feed-forward DSNN; the next section introduces the TDC algorithm.

## 2 TDC Learning Scheme

The TDC starts by mapping the input spike train/s to a series of spike trains in the network intra-layer neurons. These spike trains are called feature-domain spike trains; the feature-domain spike trains are the spatio-temporal representation of the input spike train/s generated by the TS mapping of the input spike trains. The TDC suggests reassembling of the intra-layer – here, the first layer – neurons' spike timing to build the desired output spike train; in other words, the output neurons of the DSNN coordinate the intra-layer spike timing to generate the desired spike train/s. Figure (3) depicts the TDC learning idea, and equation set (4) defines the mathematical explanation of the feature domain. In Figure (3), $DSNN_a$ represents two-layer DSNN model, and $DSNN_b$ and $DSNN_c$ are the building blocks of the network. $DSNN_b$ is a single layer DSNN and processes the input spike trains to build the feature-domain. $DSNN_c$ is also a single layer DSNN which processes both the feature-domain – $F_d$ – and input spike trains – $In$ – to generate the desired output spike train - $Out$. In the figure, $Z_d$ refers to the collection of the feature-domain and input spike trains. The TDC framework suggests that $F_d$ is the spatio-temporal representation of the input spike trains, which can be fed back to a single layer DSNN – $DSNN_{b'}$ – to reconstruct a similar representation of the input spike trains. Thus, the feature-domain spike trains are a mapping from time to space and time – here, the space dimension is represented by neurons – preserving the information carried by the input spike trains. Equation (4) describes how the feature-domain spike trains can be generated given the input spike trains. Given the objective function, the TDC proposes that feature-domain spike trains will be sparse in both time and space. The objective function is minimized by adjusting parameters of neurons in $DSNN_b$ network and it is defined by

$$\min \sum_d \sum_i \sum_n 1_n^{\text{ap},d,i} + \lambda * \sum_d \sum_i \sum_j \sum_n 1_n^{\text{ap},d,i} * 1_n^{\text{ap},d,j} \qquad (4)$$

$$\text{subject to } \sum_d \sum_m \sum_n \left| S(\text{In}_n^{d,m}, \hat{In}_n^{d,m}) \right|_0 \leq K$$

where, $d$ is the sample index of input data and $m$ is the spike train index. $1_n^{\text{ap},d,i}$ is one on the spike time, and it represents the spiking activity of $i^{th}$ spike train for $d^{th}$ sample of input signal – $In$. The objective function is minimized when the feature-domain spike trains are sparse in time and space; in the other word, each neuron of $DSNN_b$ fires in a time slot different for any other neuron of the layer. $DSNN_{b'}$ is generally built by depressing synapses; thus, its output spike timing is an assembly of its input spike trains. Given this characteristic of the $DSNN_{b'}$, the objective function defined in equation (4) looks for the sparse set of spike trains, where their spiking activities can be combined to build the most similar resemblance of the input. There other terms which might change the minimum of the objective function are $K$, $\lambda$ and number of inter-neurons; thus, the $F_d$ and model parameters – $p$ – will change as these parameters change. We will discuss later that the choice of $F_d$ will be determined by the task being run in the network.

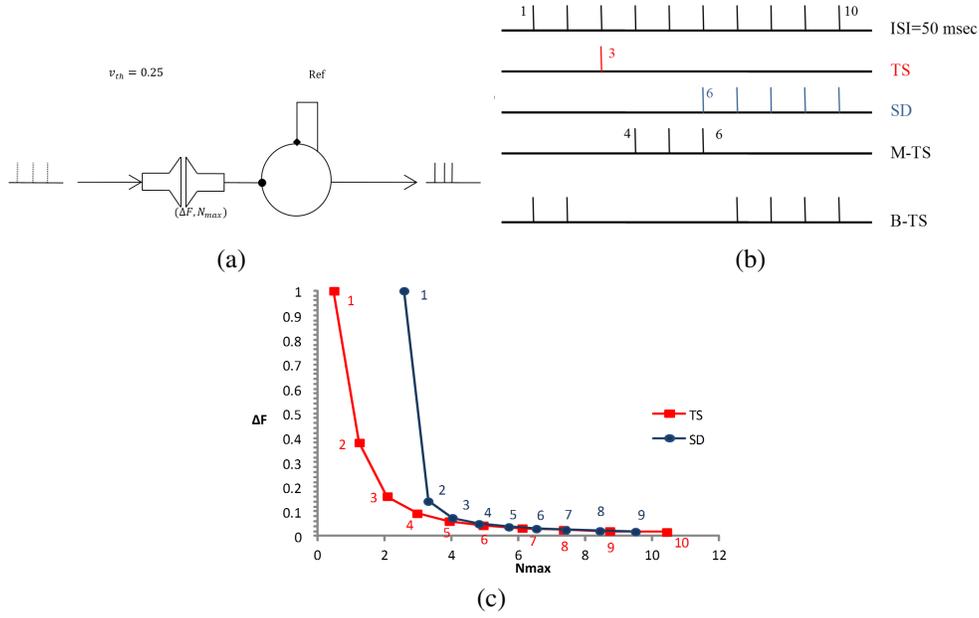

Figure 2: Temporal processing in a single synapse DSNN. (a) The motif of a single synapse DSNN, (b) A sample temporal selectivity (TS), spike activity delay (SD), multiple temporal selectivity (M-TS), and bi-modal temporal selectivity (B-TS) mappings of the input spike train, (c) Optimum parameters in the TS and SD mapping. The TS mapping corresponds to picking a specific spike of the input sequence. The M-TS mapping corresponds to picking multiple consecutive spikes of the input sequence, and the SD mapping is a specific type of M-TS. In the SD mapping, the output sequence is an M-TS mapping including all upcoming spikes of the input. The B-TS mapping is defined by a montage of multiple TS or M-TS mapping.

Figure (4.a) shows a possible solution of $Z_d$ in a two-layer DSNN with a single – or multiple – input spike train. The solution suggests dividing the processing time into non-overlapping time windows,

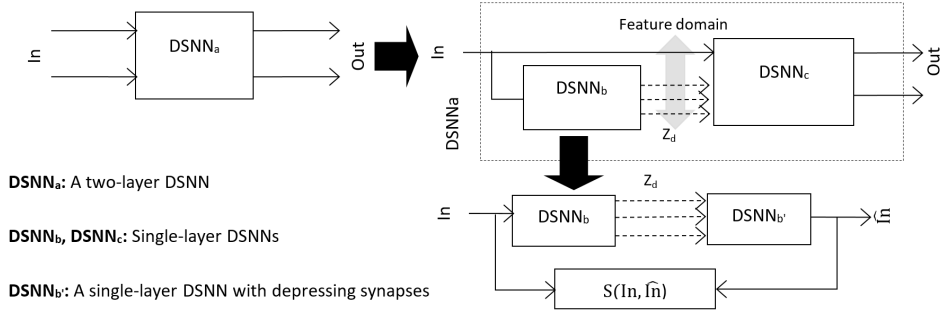

**DSNN$_a$:** A two-layer DSNN

**DSNN$_b$, DSNN$_c$:** Single-layer DSNNs

**DSNN$_{b'}$:** A single-layer DSNN with depressing synapses

Figure 3: TDC learning idea. The $DSNN_b$ network maps the input spike trains to the feature-domain spike trains ; the output network - $DSNN_c$ - receives both the input and feature-domain signals- $Zd$. The similarity measure – $S(In_i, In_j)$ – categorizes spike pairs to three classes of missing, extra, similar spikes; note that the spike similarity measure is insensitive to the spike timing jitter. For the similarity measure, extra and missing spikes generate non-zero values; for a pair of similar spike trains, the sum of non-zero values shrink to zero [21]. The feature-domain signal is a specific representation of the input signal, which satisfies the objective function defined in equation (4)

where each intra-layer neuron is trained to fire a spike – in response to input stimuli – in its time window. The window length is about the refractory period, which limits the number of spike to maximum one for each intra-layer neuron. Thus, each inter-neuron runs a TS process in response to the input stimuli. Figure (4.b) and (4.c) show two different topologies for the 2-D layer DSNN; here, we focus on the feed-forward topology. For the recurrent topology, we might assume parameters of the recurrent connections are fixed and known.

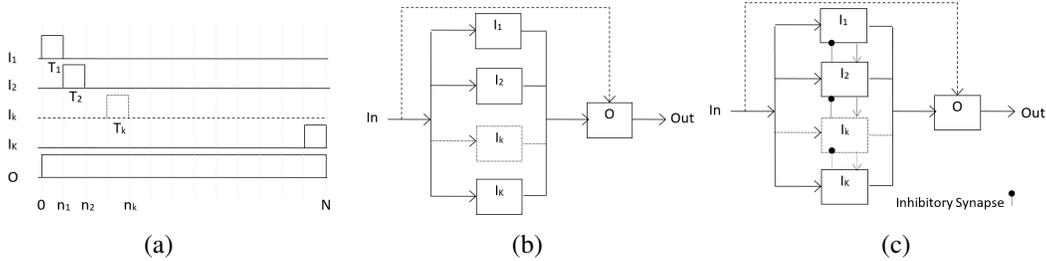

(a)  (b)  (c)

Figure 4: Non-overlapping time windows in the TDC algorithm and two-layer DSNN topology. **(a)** non-overlapping time windows for intra-layer neurons, **(b)** two-layer DSNN topology, **(c)** two-layer DSNN with recurrent connections.The connection between nodes can be inhibitory or excitatory.

The TDC framework proposes the DSNN topology and optimum spiking activity of each intra-layer neuron. The proposed spike timing is used to train each neuron's parameters; through this step, we initialize each neuron's parameters to satisfy the cost function defined in equation (4). Note that, this step of training is done without knowing the desired output spiking activity of the network. For example, the TS curve shown in figure (2.c) defines the initial parameters of different inter neurons for a spike train with 50 milliseconds ISIs. In general, when there are multiple spike trains with different spiking patterns, we can estimate each inter-neuron parameters using the training procedure discussed in the next section. Note that learning of a DSNN requires a final phase of synaptic adaptation, which forms the desired output spike train/s. In the next section, we will discuss how to run the learning in a two-layer DSNN model.

## 2.1 Learning Rule

The gradient-gain learning rule, proposed in Yousefi et al., defines a generalized framework for the synaptic adaptation in a single-layer DSNN [22]. Synaptic parameters of each synapse are adjusted by a weighted sum of its PSP gradient; the weight is defined by a gain function – $g_n$ – which is an evaluative function of the DSNN response to an input stimulus. The update rule for the synaptic parameters in the output neurons is defined by:

$$p^m \leftarrow p^m + \sum_{n=0}^{N} g_n * \nabla v_n^m / \nabla p^m \tag{5}$$

here, we assume there is only one output spike train. The $p^m$ variable refers to free synaptic parameters of $m^{th}$ synapse, and $v^m$ is the $m^{th}$ synapse PSP. Though the gradient-gain suggests the same gain function for all output synaptic connections; we condition parameter adaption in synaptic connections from input to output – the dashed line in figure (4.b) – on the spiking activity of the network inter-neurons. This conditional adaptation is a mechanism to balance synaptic strength of: a) input to output, and b) intra-layer to output synaptic connections. Under this assumption, synaptic adaptation in input to output synapses gets postponed until spiking activities in intra-layer neurons are being stabilized. The conditional adaptation idea comes from the adaptation mechanism observed in the direct synaptic pathway of Hippocampus EC $\rightarrow$ CA3 circuitry; the adaptation in this path is conditioned on dentate gyrus - DG - neurons' spiking activity [24]. The learning rule for input to output synaptic connections is defined by:

$$p^d \leftarrow p^d + \sum_{n=0}^{N} g_n * c_n * \nabla v_n^d / \nabla p^d \tag{6}$$

where, $p^d$ - $d$ stands for direct - variable is the free synaptic parameters of the input to output synaptic connection. The learning rule in equation (5) applies to parameters of the synaptic connections fed by spiking activity of inter-neurons, and equation (6) defines the learning rule in parameters of synapses being fed by the input spike trains. Equation (7) defines the conditional activation function:

$$c_n = \left| \sum_m \sum_{t \in T_c^n} ap_t^m \right|_0 \tag{7}$$

where, $T_c^n$ defines the conditional activation time window for time $n$, and $ap_t^m$ is the spiking activity of $m^{th}$ intra-layer neuron at time index $t$. For example, $T_c^n$ can be set to $T_n$ defined in figure (4.a).

TDC implies that spiking activity of each intra-layer neuron will derive the output layer to firing level. Each intra-layer neuron is trained to replicate the desired output spiking activity in its defined time window; thus, the neuron receives the output neuron gain for its defined time window. In the meantime, each neuron gets punished for spiking activities generated beyond its defined time window. The gain function for the $k^{th}$ intra- layer neuron is defined by:

$$g_n^k = \begin{cases} -1 * ap_n^k & n \notin T_k \\ g_n & n \in T_k \end{cases} \quad T_k \in \{n_{k-1} < n \leq n_k\} \tag{8a}$$

$$p^k \leftarrow p^k + \sum_{n=0}^{N} g_n^k * \nabla v_n^k / \nabla p^k \tag{8b}$$

Equations (4) and (8) define the procedure of learning in a 2-layer DSNN using TDC learning framework. Equation (4) suggests a sparse spiking activity for each inter-neuron independent of the desired output spiking activity; generally, we use the learning rule proposed in equation (5) to initialize the synaptic parameters of each inter-neuron. In the next step, equation (8) proposes how the inter-neurons' synaptic parameters to be adjusted along with other parameters of the model. Equation (8) matches the idea of TDC, where it assumes each inter-neuron contributes in constructing a portion of the desired output spike train.

Equation (8) suggests any spiking activity beyond an inter-neuron assigned time window to be punished by a negative gain – in the gradient-gain learning, a negative gain is linked to undesired extra spiking activity; this hypothesis can be applied whenever an inter-neuron spiking activity outside its assigned time leads to a negative gain. Under this assumption, the spurious spiking activity of an intra-layer neuron turning to the desired spike is preserved to facilitate the network learning process. The new gain function is defined by:

$$g_n^k = \begin{cases} \min(0, g_n) * ap_n^k & n \notin T_k \\ g_n & n \in T_k \end{cases} \tag{9}$$

Equation set (5)-(9) address the synaptic adaptation of each neuron of a two-layer DSNN. The learning process starts by adapting inter-neuron synaptic parameters given TDC learning proposal, and the main learning process starts by observing DSNN response to an input stimulus, estimating the gain function, and adjusting the network free parameters. In practice, $g_n$ is defined by evaluating the network spiking activity to input spike trains and its similarity to desired spike train. The learning framework proposed here is defined independent of $g_n$ definition, and it is applicable to different input-output spike-domain tasks. In the next section, we discuss the application of a two-layer DSNN and TDC learning algorithm in the jittered spike trains classification task.

## 3 Classification of Jittered Spike Trains

Two spike trains with a Poisson ISI distribution and the same firing rate are generated - mean firing rate of 20 Hz or average ISI of 50 milliseconds – within a time of 500 milliseconds; these spike trains are called Template A and Template B. The jittered version of the templates is generated by shifting each spike of the templates with a random time drawn from a Gaussian distribution with zero mean and 4 milliseconds standard deviation. The classification task is to determine the template – either A or B – from which the jittered spike train is being generated. The templates share the same mean firing rate; thus, their distinct features are spiking timing and ISIs variability. We generate 100 jittered versions of each template. We use 65 samples of each template for the training step, and we use the remaining 35 samples of each template to check the model classification performance [21].

### 3.1 DSNN Topology and Gain Function

The DSNN has two output neurons; each neuron is trained to fire a spike at an earlier time in response to its matching template. The learning objective function in response to a sample of the Template A is defined by:

$$max_P \; t_{AP}^B - t_{AP}^A \cong min_P \sum_{n=t_{exp}^A}^{t_{AP}^A} E_n^{ap,A} + \sum_{n=0}^{t_{AP}^A} E_n^{non-ap,B} \qquad (10)$$

where, variable $t_{AP}^A$ denotes the time of the first spike in the matching output neuron – neuron A, and $t_{exp}^A$ is the expected time for the first spike in the same output neuron. Variable $E_n^{ap,A}$ is the cost to generate a spike at time $n$ for output neuron A, and variable $E_n^{non-ap,B}$ is the cost for the non-spike time period of output neuron B. Variable $t_{exp}^A$ is the moving average of $t_{AP}^A$, which determines the learning stop criterion. Variable $P$ defines the network free parameters including both output and intra-layer neurons' synaptic parameters. The cost function is maximized when the matching neuron fires much earlier than the competing neuron; in the other word, it is minimized when the cost of firing for the matching neuron is minimized and the cost of firing for the competing neuron is maximized. Using equation (10), the gain function for the output neurons in response to a sample of Template A is defined by:

$$g^A(n) = \begin{cases} 1 & t_{exp}^A < n \leq t_{AP}^A \\ 0 & \text{Otherwise} \end{cases} \qquad (11a)$$

$$g^B(n) = \begin{cases} -1 & t_{AP}^B < n \leq t_{AP}^A \\ 0 & \text{Otherwise} \end{cases} \qquad (11b)$$

$$t_{exp}^A = (1-\alpha) * t_{exp}^A + \alpha * t_{AP}^A \qquad 0 < \alpha < 1 \qquad (11c)$$

Equation (11.a) suggests a positive gain for output neuron A in the time period $t_{exp}^A$ to $t_{AP}^A$ – note that $t_{AP}^A$ becomes 500 milliseconds if neuron A fails to fire a spike; thus, the gain function pushes the neuron A to fire a spike in the earliest time. The gain for the competing neuron – neuron B – penalizes any spiking activity that happens before the first spike in neuron A. Thus, the gain function becomes zero – a zero gain means observing a desired response in the network, if neuron A fires ahead of neuron B in response to a sample of Template A and vice versa.

Table 1: Performance of two-layer DSNN and alternative models in the jittered spike classification task

| Network Type | Accuracy (training/test dataset) % |
|---|---|
| Two-layer DSNN | 92.5/91.6 |
| Two-layer DSNN with static synapses | 33.7/30.8 |
| Single-layer DSNN | 70.3/67.8 |

Using the TDC proposal for the DSNN topology, each output neuron receives spiking activities of 10 intra-layer neurons plus a direct synaptic connection from the input spike train. For the network, all intra-layer neurons receive the same input spike train. The whole processing time is partitioned to 50-millisecond windows, and the time window for $k^{th}$ intra-layer neuron is defined by $T_k = [\ (k-1)*50 \quad k*50\ ]$ milliseconds. Note that the TDC suggests a larger number of intra-layer neurons with a shorter time window, the 50-millisecond window is chosen based on the average ISIs of the input stimulus and the complexity of this classification task.

The learning rule for the DSNN is defined by the equation set (11) plus equations (5-7) and (9). The initial value for the synaptic parameters of intra-layer neurons is defined by the TS curve shown in figure (2.c). Note that inter-neurons' synaptic parameters can be updated using the gradient-gain learning algorithm; each inter-neuron is trained to fire a spike its time window in response to all training data samples – both Template A and B. The synaptic parameters of direct input to output path is set by a low facilitation factor with a sub-threshold synaptic strength – a synapse with a low facilitation factor is a potentiating synapse, which increases its vesicle release in response to a sequence of spikes; other output neurons synaptic connections are initialized by a high facilitation factor with supra-threshold strength – a synapse with a high facilitation factor is a depressing synapse, releasing its whole vesicle pool in response to the first impinging spike. Figure (5) depicts the DSNN topology plus synaptic parameters in a trained DSNN.

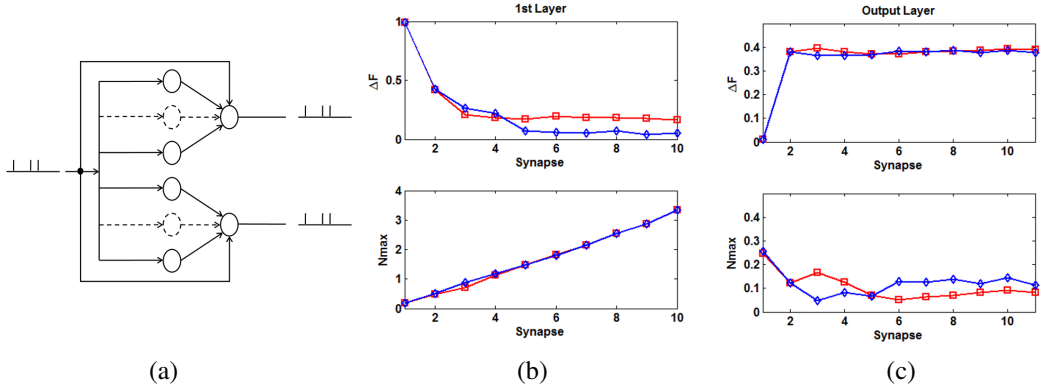

Figure 5: Two-layer DSNN topology for the jittered spike train classifier and synaptic parameters of the trained DSNN. **(a)** DSNN model, **(b)** Synaptic parameters in the first layer **(c)** Synaptic parameters in the output layer - the first synapse in output layer belongs to the direct synaptic path. A synapse with a lower index belongs to an intra-layer neuron with an earlier time window – the network firing threshold - $v_{th}$ - is set to 0.25

## 3.2 Performance Analysis

We trained the two-layer DSNN plus two other variant models for this task. The other models are a two-layer SNN with static synapses - for the static synapses, the only free parameter is synaptic weight, and a single-layer DSNN. Table 1 shows the classification accuracy for all three networks. Here, the accuracy is defined by the number of correctly classified trains divided by the total number of classification being run – 700 trial here.

Neither the two-layer SNN with static synapses nor the single-layer DSNN shows a comparable performance in comparison to the two-layer DSNN. The time integration capacity of a static synapse is limited to its PSP time constant, which is only in the range of 10 to 100 milliseconds. The discriminant feature of the input stimulus is embedded in its spike timing spanning for a couple of hundred milliseconds; therefore, the static synapse SNN fails to reach a comparable accuracy. The single-layer DSNN lacks the higher dimension of inputs feeding the output neurons of the two-layer DSNN; thus, it requires a fine tuning of its synaptic parameters to perform the same task. The synaptic response of a single-layer DSNN becomes sensitive to spike jitter reducing its classification accuracy - a similar phenomenon to what being observed in the tail of the TS curve. In contrast, the two-layer DSNN shows a high accuracy in performing the task. In fact, its performance can be enhanced by using a higher number of inter-neurons. The classification task is distributed between inter-neurons and output layer neurons; as a result, the output layer shows a less sensitivity to the jitter which leads to a better performance.

Figure (6) shows a trained DSNN response to one sample of each template. The intra-layer synapses of the DSNN are trained to build a task-dependent sparse representation of the input stimulus; at the same time, the output neurons are trained to choose one of the intra-layer spike timing to correctly classify template A and B. Note that static synapses are incapable of performing the TS process; as a result, the temporal processing capability of the network is limited and does not grow by adding layers.

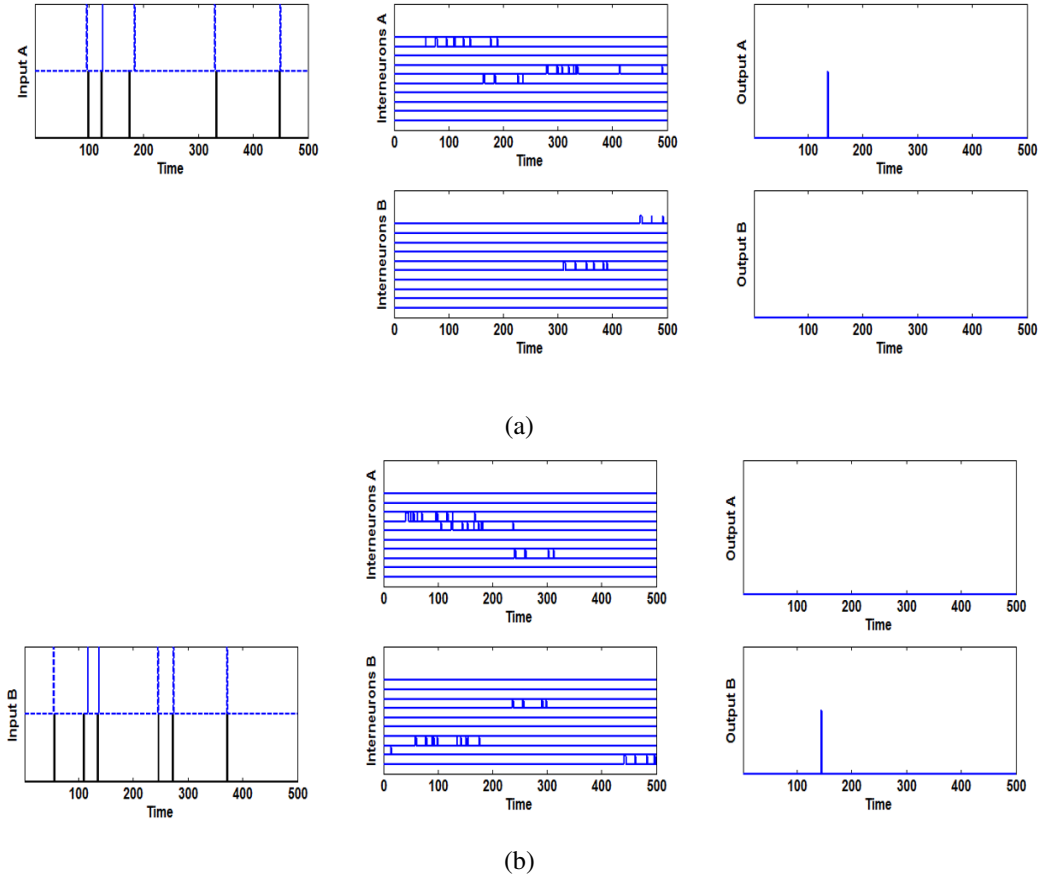

Figure 6: Two-layer DSNN response to a sample of both templates. **(a)** DSNN neurons' spiking activity in response to a sample of Template A. The black lines show the Template A and B spike trains, and the blue lines show the jittered spike trains. **(b)** DSNN neurons' spiking activity in response to a sample of Template B. On the left, we have the input spike train – jittered sample in blue - and its template – in black. The middle figures show spiking activity of inter-neurons; note that inter-neurons are not necessarily sparse and might fire outside of their defined time window.

## 4  Discussion

The jittered spike classification task is a spike-domain processing task, designed to assess the capability of a neural network to process neural information carried through precise spike timing. We utilized DSNN – a brain-inspired neural network – along with TDC learning to perform the jittered spike classification task; the proposed model reaches more than 90% accuracy in classifying different spike train templates. The model performance is 50% better than the same network topology with static synapses. This is a general caveat of static models and it requires changing the network topology including recurrent connections; whereas, the DSNN has the capability of processing dynamic signals without changing the network topology. Beyond the high accuracy of classification achieved using the DSNN, we also demonstrated that network capability to utilize temporal features of the input to decide in a minimal time. We also demonstrated how the network gains from a successive layer of temporal processing to boost its performance; for instance, 2-layer DSNN shows 20% improvement comparing to a single layer DSNN.

In the jittered spike classification problem, we assigned one output neuron to each template – template A and B. It is possible to change the network topology – and its learning objective function – to utilize the TDC idea for training this new topology. For instance, we might have only single output neuron and this neuron will fire in different time slots depending on the type of input template. Here, we proposed the TDC learning for a multiple-input-single-output DSNN – note that in the jitter classification task, we train two separate single-input-single-output DSNNs. The TDC is scalable for the general class of multiple-input-multiple-output DSNNs. There are multiple solutions for these network; the first solution is similar to what has been utilized in the jitter classification task; here, each output neuron receives a separate set of inter-neurons proposed by the TDC. The second solution includes a DSNN with only one set of inter-neurons shared between all output neurons; for this topology, the gain of inter-neurons corresponds to the majority of proposed gain by the output neurons. For example, if the proposed gains – gn for time n – for an inter-neuron are 1, 1, 0, and -1; the proposal gain for the inter-neuron becomes 1. The proposal gain can be also defined by the sign of proposed gains sum. Using the TDC, we increase the depth of the network and propose a learning rule for its deeper layers. It is trivial to show that a 2-layer DSNN attains a higher computation power than its single-layer counterpart, but the extent of this improvement requires further theoretical and experimental analysis. In developing this learning framework, we hypothesize that the deeper layer neurons have a sparse activity in both time and space – with space represented by each neuron. Though the sparseness property has been observed in neural circuits, we would need to run the proposed framework in more complex tasks to validate the TDC methodology performance and stability. In general, we proposed the TDC from a machine learning perspective. Yet, we could also analyze the learning mechanism and its performance from the statistical point of view. The spike time could be described by as a point process, and we could use point-process theories to examine both the learning procedure proposed here and its result. Furthermore, we could use the state-space framework to describe dynamics of DSNN synapses. New theories to analysis point process along with dynamical model could be applied here for further assessment of the proposed framework in future studies. In developing this framework, the work depended heavily on the analysis of the synaptic response and its temporal dynamics. We also utilized the concepts of sparse representation by incorporating neural circuitries' spiking properties. The idea of TDC proposed here can be extended to a deeper layer of a network, where each layer is trained to reduce the complexity of the task being processed in the next layer. Deep neural network models have received a significant attention recently. The DSNN has interesting properties particularly since it is built upon both STP and LTP characteristics of neural circuitries and it is capable of processing temporal signals. The TDC idea proposed here can be extended in deep layers of a network and combined with more complex topologies to solve more complex engineering and neuroscience related problems.

## 5  Conclusion

We proposed TDC learning for training a two-layer DSNN model. The learning proposed here is capable of simultaneously adapting LTP and STP parameters of synaptic connections of the network. The framework can be extended in learning multiple-input-multiple-output DSNN with multiple layers of neurons and synaptic connections. We used DSNN to classify the jittered spike trains problem proposed to examine the performance of multi-layer recurrent neural networks; we showed more than 92% accuracy in this task only using a two-layer DSNN and TDC learning. Some next steps are: a) to provide further theoretical and experimental analysis of the learning framework, and b) using DSNN to predict place cell spiking activity of a rat moving in the 2D maze.

## Acknowledgment

We thank Dr. Behzad Nazari (Department of Electrical and Computer Engineering, Isfahan University of Technology), Dr. Angelique C. Paulk (Harvard Medical School, Department of Neurosurgery) and Yalda Amidi (PhD Student in Department of Electrical and Computer Engineering, Isfahan University of Technology) for their comments and help that greatly improved the manuscript.